\RequirePackage{fix-cm}
\documentclass[smallextended]{svjour3}       
\usepackage[utf8]{inputenc}
\usepackage{graphicx}

\usepackage{xspace}
\usepackage{amssymb}
\usepackage{amsmath}
\usepackage{mathtools}
\usepackage{booktabs,multirow}
\usepackage{verbatim}
\usepackage{datatool}
\usepackage{graphics}
\usepackage[hidelinks]{hyperref}
\newcommand{\citet}{\cite}
\usepackage[showboxes]{textpos}
\usepackage{subfig}
\usepackage{tikz}
\usetikzlibrary{arrows,shapes}
\usepackage{comment}
\usepackage{tabularx}

\newcommand{\bftab}{\fontseries{b}\selectfont} 

\newcommand{\mltc}[1]{\multicolumn{2}{c}{#1}}

\newcommand{\gamename}{\textsc}

\newcommand{\GHex}{\textsc{Generalized Hex}\xspace}
\newcommand{\complexityclass}[1]{\textsf{\footnotesize #1}\xspace}
\newcommand{\pspace}{\complexityclass{PSPACE}}

\newcommand{\DYS}{\emph{DYS}\xspace}

\newcommand{\ENCP}{\emph{COR+}\xspace}
\newcommand{\EncodingName}{Corrective\xspace}
\newcommand{\EncodingNameP}{Corrective+\xspace}

\newcommand{\white}{{\textsc{W}} }
\newcommand{\black}{{\textsc{B}} }
\newcommand{\board}{{\mathsf{board}}}
\newcommand{\timee}{\mathsf{time}}

\newcommand{\win}{\mathsf{win}}
\newcommand{\moveL}{\mathsf{move}}
\newcommand{\depth}{\mathsf{F}}
\newcommand{\size}{\mathsf{N}}

\newcommand{\relmiddle}[1]{\mathrel{}\middle#1\mathrel{}}
\DeclarePairedDelimiter\ceil{\lceil}{\rceil}

\usepackage{acronym}
\acrodef{GTTT}{Generalized Tic-tac-toe}
\acrodef{MCTS}{Monte Carlo Tree Search}
\acrodef{PNS}{Proof Number Search}
\acrodef{QBF}{Quantified Boolean Formula}
\acrodefplural{QBF}{Quantified Boolean Formulas}
\acrodef{CNF}{Conjunctive Normal Form}
\acrodef{SAT}{Satisfiability Problem}

\definecolor{light-gray}{gray}{0.92}
\tikzset{
  hav-white/.style={draw,circle,fill=light-gray,minimum size=3.5mm,inner sep=0,text=black,font=\scriptsize},
  hav-black/.style={draw,circle,fill=black,minimum size=3.5mm,inner sep=0,text=white,font=\scriptsize},
  hav-noone/.style={text=black,font=\scriptsize},
  hav-empty/.style={shape=regular polygon,regular polygon sides=6,draw,minimum size=5.85mm,inner sep=0,outer sep=0,font=\scriptsize},
}
\newcommand{\havcoordinate}[4]{
  \foreach \i in {#1,...,#2} {
    \foreach \j in {#3,...,#4} {
      \coordinate (\i-\j) at ({(\i+1) * 1.732},{(\j)});
    }
  }
  \foreach \i in {#1,...,#2} {
    \foreach \j in {#3,...,#4} {
      \coordinate (\i--\j) at ({(\i+1+1/2) * 1.732},{(\j+1/2)});
    }
  }
}

\newcommand{\vaw}{\begin{tikzpicture}\node[hav-white,font=\normalsize] {$a$};\end{tikzpicture}}
\newcommand{\veb}{\begin{tikzpicture}\node[hav-black,font=\normalsize] {$e$};\end{tikzpicture}}
\newcommand{\vib}{\begin{tikzpicture}\node[hav-black,font=\normalsize] {$i$};\end{tikzpicture}}

\smartqed  
\usepackage{graphicx}
\journalname{}
\begin{document}

\title{Positional Games and QBF: \\ A Polished Encoding}

\titlerunning{Games and QBF}        

\author{Valentin Mayer-Eichberger \and
Abdallah Saffidine}
\authorrunning{V. Mayer-Eichberger A. Saffidine}

\institute{Universit\"at Potsdam, Germany \and
University of New South Wales, Sydney, Australia\\\email{abdallah.saffidine@gmail.com} (corresponding author)}

\date{Submitted 22. Feb. 2021 \footnote{This is a technical report extending our SAT 2020 paper \emph{Positional games and QBF: the corrective encoding} \cite{Saffidine20}}.}

\maketitle
\begin{abstract}

Positional games are a mathematical class of two-player games comprising
Tic-tac-toe and its generalizations. We propose a novel encoding of these games
into \acp{QBF} such that a game instance admits a winning strategy for the first player 
if and only if the corresponding formula is true. Our approach improves over
previous \ac{QBF} encodings of games in multiple ways. 
First, it is generic and lets us encode other positional games, such as Hex. 
Second, the structural properties of positional games, together with careful
treatment of illegal moves, let us generate more compact instances that can be
solved faster by state-of-the-art \ac{QBF} solvers. We establish the latter
fact through extensive experiments.
Finally, the compactness of our new encoding makes it feasible to translate realistic
game problems.
We identify a few such problems of historical significance and put them forward
to the \ac{QBF} community as milestones of increasing difficulty.

\end{abstract}

\section{Introduction}

Solving larger and larger computational challenges requires scientific progress in various areas, from hardware performance to algorithmic efficiency through aggressive compiler optimizations.
Our main focus in this article is the area of logical encodings.
By logical encoding, we refer to the process of expressing a mathematically well-defined problem or class of problems in a restricted logical formalism such that answering the logical query corresponds to answering the mathematical problem.
To be more specific, for any instance of a positional game (a class of two-player strategy games that includes \gamename{Tic-tac-toe} and \gamename{Hex}), we construct a \emph{high-quality} \ac{QBF} that is true if and only if the Black player wins the input position.

Our line of work has three objectives:
understand better how to write high-quality logical encodings of Artificial Intelligence problems;
stimulate the development of logic solvers by providing challenging benchmark instances; and
contribute to bridging the gap between the generic problem-solving approach based on logic and domain-specific algorithms.
We progress in all three directions by developing an encoding of the AI problem (positional games) into the logical formalism (\ac{QBF}), applying our new encoding to \gamename{Hex}, \gamename{Generalized Tic-tac-toe}, and other games, and running extensive experimental analyses involving state-of-the-art \ac{QBF} solvers and measuring their performance on our encoding as well as on a related encoding described previously by other authors~\cite{DiptaramaYS2016}.

Unlike the discipline of SAT Solving \cite{JarvisaloLBRS2012} and the development of artificial game solvers and players~\cite{SchaefferBBKMLLS2007,SilverHSALGLSKG2018}, the study of encodings in \ac{QBF} is relatively underdeveloped, especially as far as strategy games are concerned~\cite{Seidl20}.
This is surprising given the well-understood conceptual connections between \acp{QBF} and 2-player strategy games ranging from \emph{Game Semantics} to \pspace-hardness.
In 2003, this observation prompted Toby Walsh to challenge the \ac{QBF} community to solve \gamename{Connect4} on a $7\times6$ board~\cite{Walsh2003}, a problem solved by game-specific software since 1988~\cite{Allis1988}.
Subsequently, there was activity to tackle this problem, but with little practical success.
The \ac{QBF} community was young and small, with limited knowledge in clausal encodings and no strong solvers available.

The first concrete and implemented encoding of a game to \ac{QBF} is the work by Gent and Rowley that presents a translation from \gamename{Connect4} to \ac{QBF} \cite{GentA2003}.
Building upon Gent's encoding \cite{Gomes05} presents a \ac{QBF} encoding of an \emph{Evader/Pursuer} game that resembles simpler chess-like endgames on boards of size $4\times4$ and $8\times8$.
Both papers analyse problems that are not positional games, but the authors do report similar challenges in the construction of \ac{QBF} formulas.

Various encodings of any given mathematical problem may exist, and in practice, it is important to be able to find ``good encodings''. 
There are various reasonable axes of comparison between encodings.
\begin{description}
\item[Input Generality:] covering a larger class of mathematical problems is better.
\item[Output Specificity:] targeting a more restricted logical language is better.
\item[Readability:] being easier for humans to understand is better.
\item[Size:] leading to smaller formulas (in terms of the number of variables, clauses, quantifier blocks, etc.) is better.
\item[Performance:] leading to formulas that state-of-the-art solvers can address faster is better.
\end{description}
These dimensions of analysis sometimes overlap, but they do not coincide.
After having described our new \ENCP Encoding of positional games in \ac{QBF}, we will present evidence that \ENCP scores well along all these axes, at the least when compared to related work designed for the \ac{GTTT} fragment of the class we address: positional games.

In a \emph{positional game}~\cite{HalesJ1963,BonnetGLRS2017}, two players alternately claim unoccupied elements of the board of the game.
The goal of a player is to claim a set of elements that form a winning set, and/or to prevent the other player from doing so.
\gamename{Tic-tac-toe}, and its competitive variant played on a $15\times 15$ board, \gamename{Gomoku}, as well as \gamename{Hex} are the most well-known positional games.
When the size of the board is not fixed, the decision problem, whether the first player has a winning strategy from a given position in the game is \pspace-complete for many such games.
The first result was established for \GHex, a variant played on an arbitrary graph~\cite{EvenT1976}.
Reisch \cite{Reisch1980} soon followed up with results for \gamename{Gomoku}~\cite{Reisch1980} and \gamename{Hex} played on a board~\cite{Reisch1981}.


Recent work on the classical and parameterized computational complexity of positional games provides us with elegant first-order logic formulations of such domains~\cite{BonnetJS2016,BonnetGLRS2017}.
We draw inspiration from this approach and introduce a practical implementation for such games into \ac{QBF}.
We believe that Positional Games exhibit a class of games large enough to include diverse and interesting games and benchmarks yet allow for a specific encoding exploiting structural properties.

Our contributions are as follows:
(1) we introduce the \emph{\EncodingNameP Encoding}: a generic translation of positional games into \ac{QBF};
(2) we identify a few positional games of historical significance and put them forward to the \ac{QBF} community as milestones of increasing difficulty;
(3) we demonstrate on previously published benchmark instances that our encoding leads to more compact instances that can be solved faster by state-of-the-art \ac{QBF} solvers;
(4) we establish that the \EncodingNameP Encoding enables \ac{QBF} solving of realistic small-scale puzzles of interest to human players.

After a formal introduction to \ac{QBF} and positional games (Section~\ref{sec:preliminaries}), we describe the contributed translation of positional games into \acp{QBF} (Section \ref{sec:encoding}).
We then describe the selected benchmark game problems, including the proposed milestones (Section \ref{sec:instances}), before experimentally evaluating the quality of our encoding and comparing it to previous work (Section~\ref{sec:analysis}).
We conclude with a discussion contrasting our encoding with related work (Section \ref{related-works}).

\section{Preliminaries on \ac{QBF} and Positional Games}
\label{sec:preliminaries}

We assume a finite set of propositional variables $X$. A literal is a variable $x$
or its negation $\neg x$. A clause is a disjunction of literals. 
A \ac{CNF} formula is a conjunction of clauses. 
An assignment of the variables is a mapping $\tau:X\rightarrow \{\bot,\top\}$.
A literal $x$ (resp. $\neg x$) is satisfied by the assignment $\tau$ if
$\tau(x)=\top$ (resp. $\tau(x)=\bot$).
A clause is satisfied by $\tau$ if at least one of the literals is satisfied.
A \ac{CNF} formula is satisfied if all the clauses are satisfied. 

A QBF formula (in \emph{Prenex}-CNF) is a sequence of alternating blocks of
existential ($\exists$) and universal ($\forall$) quantifiers over the
propositional variables followed by a \ac{CNF} formula. 
A \ac{QBF} formula may be interpreted as a game where an $\exists$ and
$\forall$ player take turns building a variable assignment $\tau$ selecting the
variables in the order of the quantifier prefix. 
The objective of $\exists$ (resp. $\forall$) is that $\tau$ satisfies (resp.
falsifies) the formula. 


\emph{Positional games} are played by two players on a hypergraph $G=(V,E)$.
The vertex set $V$ indicates the available positions, while each hyperedge $e\in E$ denotes a winning configuration.
For some games, the hyperedges are implicitly defined instead of being explicitly part of the input.
The two players alternatively claim unclaimed vertices of $V$ until either all elements are claimed, or one player wins.
A \emph{position} in a positional game is an allocation of vertices to the players who have already claimed these vertices.
The \emph{empty position} is the position where no vertex is allocated to a player.
The notion of winning depends on the game type.
In a \emph{Maker-Maker game}, the first player to claim all vertices of some hyperedge $e \in E$ wins.
In a \emph{Maker-Breaker game}, the first player (\emph{Maker}) wins if she claims all vertices of some hyperedge $e\in E$.
If the game ends and player 1 has not won, then the second player (\emph{Breaker}) wins.
The class of $(p, q)$-positional games is defined similarly to that of positional games, except that on the first move, Player 1 claims $q$ vertices and then each move after the first, a player claims $p$ vertices instead of 1.
A \emph{winning strategy} for player 1 is a move for player 1 such that for all moves of player 2 there exists a move of player 1\dots such that player 1 wins.

To illustrate these concepts, Figure~\ref{fig:GamesDef} displays a position from a well-known Maker-Maker game, \gamename{Tic-tac-toe}, and a position from a Maker-Breaker game, \gamename{Hex}.
Although the rules of \gamename{Hex} are typically stated as Player 1 trying to create a path from top left to bottom right and Player 2 trying to connect the top right to bottom left, the objective of Player 2 is equivalent to preventing Player 1 from connecting their edges~\cite{Maarup2005}.
Therefore, \gamename{Hex} can be seen as a Maker-Breaker positional game.

\newcommand{\scala}{1.0}
\begin{figure}
  \subfloat[A game of \gamename{Tic-tac-toe} and its winning configurations: the set of aligned triples.]{
    \begin{tabular}{@{\quad}c@{\quad}}
    \begin{tikzpicture}[>=stealth',scale=\scala,every node/.style={scale=2*\scala}]
      \draw[thick] (1.5, 0.7) to (1.5, 3.3); \draw[thick] (2.5, 0.7) to (2.5, 3.3);
      \draw[thick] (0.7, 1.5) to (3.3, 1.5); \draw[thick] (0.7, 2.5) to (3.3, 2.5);
      \node[hav-white] at (1, 3) {$a$}; \node[hav-noone] at (2, 3) {$b$}; \node[hav-noone] at (3, 3) {$c$};
      \node[hav-noone] at (1, 2) {$d$}; \node[hav-black] at (2, 2) {$e$}; \node[hav-noone] at (3, 2) {$f$};
      \node[hav-noone] at (1, 1) {$g$}; \node[hav-noone] at (2, 1) {$h$}; \node[hav-black] at (3, 1) {$i$};
    \end{tikzpicture}\\[3mm]
  $\begin{aligned}
    & \{\vaw,b,c\},\{d,\veb,f\},\{g,h,\vib\},\\
    & \underbrace{\phantom{\,}\dots,\{\vaw,\veb,\vib\},\{c,\veb,g\}\phantom{\qquad}}_{\text{Winning sets}}
  \end{aligned}$
  \end{tabular}}
  \hfill
  \subfloat[A game of \gamename{Hex} and its winning configurations for Black (Player 1): the set of paths from the top left edge to the bottom right edge.]{
  \begin{tabular}{@{\quad}c@{\quad}}
  \begin{tikzpicture}[>=stealth',scale=\scala,every node/.style={scale=2*\scala}]
  \havcoordinate{0}{5}{0}{5}
  \node[hav-white] at ([shift={(0,-0.3)}]1--1) {}; \node[hav-black] at ([shift={(0,0.3)}]1--3) {};
  \node[hav-black] at ([shift={(0,-0.3)}]3--1) {}; \node[hav-white] at ([shift={(0,0.3)}]3--3) {};
  \node[hav-empty] at (1--2) {};
  \node[hav-empty] at  (2-2) {$d$};
  \node[hav-empty] at  (2-3) {$b$};
  \node[hav-empty] at (2--1) {$g$};
  \node[hav-empty] at (2--2) {};
  \node[hav-empty] at (2--3) {$c$};
  \node[hav-empty] at  (3-2) {$h$};
  \node[hav-empty] at  (3-3) {$f$};
  \node[hav-empty] at (3--2) {};
  \node[hav-white] at (1--2) {$a$};
  \node[hav-black] at (2--2) {$e$};
  \node[hav-black] at (3--2) {$i$};
  \end{tikzpicture}\\[2mm]
$  \begin{aligned}
    & \{\vaw,d,g\},\{\vaw,d,\veb,h\},\{\vaw,d,\veb,f,\vib\},\\
    & \underbrace{\phantom{\,}\{b,d,g\}, \{b,\veb,g\},\dots,\{c,f,\vib\}\phantom{\qquad}}_{\text{Winning sets}}
  \end{aligned}$
  \end{tabular}}
  \caption[fragile]{Two positional games played on the same vertices: $a$--$i$ where vertex $a$ has been claimed by Player 2 and vertices $e$ and $i$ have been claimed by Player 1.}
  \label{fig:GamesDef}
\end{figure}


\section{The \EncodingNameP Encoding}
\label{sec:encoding}

Producing a well-behaved \ac{QBF} formula is a challenging task, and we have
gone through many iterations to craft an encoding that works well in practice.
The version of the encoding presented here is an improvement and simplification
from the \EncodingName encoding in \cite{Saffidine20}.

We present the \EncodingNameP encoding (\ENCP) in the following way:
First, we define positional games formally.
Then, we define the set of variables used in the encoding.
The quantifier prefix is followed by a detailed explanation of each type of clauses.

\subsection{Definition of Positional Games}

A positional game is a tuple $\prod = \langle T_\black, T_\white, \depth ,
\size, E_\black, E_\white \rangle $ consisting of:

\begin{itemize}
\item Disjoint sets $T_\black$ and $T_\white$ of time points in which Black and White make moves.
  We denote $T=T_\black \cup T_\white$ as the set of all time points that range from $1$ to $\depth$.
  For example, in a positional game where black starts and $p=q=1$, $T_\black$ contains all odd and $T_\white$ all even numbers of $T$.
\item $\depth\in\mathbb{N}$ the depth (or length) of the game.
\item A set of vertices $V=\{1 \ldots \size\}$ and two sets of hyperedges $E_\black$ and $E_\white$ of winning configurations for Black and White, respectively.
\end{itemize}

The remainder of this section defines a translation of a positional game configuration $\prod$ into a prenex \ac{QBF} in CNF.
For this, we introduce variables defined in the following table.
For readability, we use a function-style notation instead of variable subscripts.

\subsection{Encoding}

Let $A$ denote the set of the two players $\{\black,\white\}$.
We introduce variables as follows:

\begin{table}[h]
\begin{tabularx}{\textwidth}{lX}
  \toprule
  Variable & Description \\
  \midrule
  $\timee(t)$     & The game is not finished at time point $t \in T$ \\
  $\board(a,v,t)$ & Player $a \in A$ owns vertex $v \in V$ at time point $t \in T$ \\
  $\win(e)$       & Black is winning with hyperedge $e\in E_\black$ \\
  $\moveL(i,t)$   & At time $t$ the player chooses the $i$th bit,  $0 \leq i < \lceil \log_2(\size) \rceil$, $t\!\in\!T$ \\
  \bottomrule
\end{tabularx}
\end{table}

The choices of selecting a vertex for both Black and White are encoded \emph{logarithmically}, so there is no single propositional variable identifying a vertex.
Vertices range from $1$ to $\size$, and there is a variable per bit of the representation of this index.
Only when all bits are known the corresponding vertex can be implied.
This uses only logarithmically many variables but the same number of clauses as a
variable per vertex encoding.
However, here, this encoding has another great benefit: $\moveL$ encodes the
choices of White in a way that prevents the universal player from falsifying
the formula by breaking the rules of the game. By design of the variables,
White cannot choose too many vertices.

\paragraph{Quantification.}

First, we specify the quantifier prefix of our encoding.
Starting with $1$ being the outermost level, we introduce in the order of the time points $t = \{1 \ldots F\} $ a level of quantifier blocks as follows:

\begin{align*}
  \begin{split}
    & \exists \timee(t)  \\
      \text{ if $t \in T_W$ }, 0 \leq i < \lceil \log_2(\size) \rceil :\;\;\;  & \forall \moveL(i, t) \\
      \text{ if $t \in T_B$ }, 0 \leq i < \lceil \log_2(\size) \rceil :\;\;\;  & \exists \moveL(i, t) \\
      \text{ for } v \in V, a \in A :\;\;\;  & \exists \board(a, v, t) \\
  \end{split}
\end{align*}

On the innermost level we have
\begin{align*}
    e \in E_\black \;\;\;  & \exists \win(e)
\end{align*}

In the remainder of this section, we list the clauses that make the body of the
generated \ac{QBF} instance. This body is constituted of sets of clauses
encoding different aspects of the game. The encoding is almost entirely
symmetric for both players, apart from clauses that specify the interaction of
the universal variables.

\paragraph{Time Handling.}
Variable $\timee(t)$ is true if the game is not over at time point $t$.
Once the game is finished, it remains so until the final time point.
\begin{align}
  \{\neg \timee(t) \vee \timee(t-1) & \mid t \in T \}\label{eq:enc14-1}
\end{align}

\paragraph{Structure of the board.}
The initial board is empty via unit clauses \eqref{eq:enc14-2}.
Clauses \eqref{eq:enc14-3} force that both players cannot own the same vertex.
One fundamental property of positional games is that claimed vertices never change owner.
This basic property is captured in clause \eqref{eq:enc14-4}.
Note that these two clauses are independent of the $\timee$ variable and also act when the game is over.
Once a vertex is claimed and $\board(a,v,t)$ is true, the implication chain in \eqref{eq:enc14-4} sets all board variables for that vertex in the future, in particular the last one $\board(a,v,\depth)$.

\begin{align}
  \{\neg \board(a,v,0) & \mid a\in A, v \in V \}\label{eq:enc14-2}\\
  \{\neg \board(\black,v,t) \vee \neg \board(\white,v,t) & \mid v \in V, t \in T \}\label{eq:enc14-3} \\
  \{\neg \board(a,v,t-1) \vee \board(a,v,t)  & \mid a \in A, v \in V, t \in T \}\label{eq:enc14-4}
\end{align}

\paragraph{Frame axiom.}

Independent of time, clause~\eqref{eq:enc14-5} forces in time points $t$ where the opponent to $a$ takes a turn, all unclaimed vertices by $s$ will be unclaimed in the next time point.
When the game is over, $\timee(t)$ is false, all empty positions are propagated through to the final time point via clause~\eqref{eq:enc14-6}.
\begin{align}
    \{\board(a,v,t-1) \vee \neg \board(a,v,t) & \mid a \in A, v \in V,  t \in T\setminus
    T_a \}  \label{eq:enc14-5} \\
    \{\timee(t) \vee \board(a,v,t-1) \vee \neg\board(a,v,t) & \mid a \in A, v
    \in V, t \in T \}\label{eq:enc14-6}
\end{align}

\paragraph{White's Choice.}
The universal variables interact crucially with the rest of the encoding.
First, they need to force a vertex for white.
Second, we need to prevent White from making illegal moves.
To avoid that White choosing too many vertices, we encode the move logarithmically through variables $\moveL(i,t)$.
By design of the variables, White cannot claim too many vertices.
Moreover, these variables only force a move of White in case the game is still running and the vertex has not been claimed by Black already.

For the logarithmic encoding, we introduce the following notation.
Let $L_1(v) \subseteq [0, \ceil{\log_2(\size)})$ denote the set of indices that are $1$ in the binary representation of $v$, likewise $L_0(v)$ denotes the set of indices where the bit is $0$.
The following equality holds: $v = \sum_{j\in L_1(v)} 2^j$.
For example, for $13=1101|_{2}$ the respective sets are $L_1(13)=\{0,2,3\}$ and $L_0(13)=\{1\}$.

\begin{align}
  \begin{split}
    & \left \{ \bigvee_{i\in L_1(v)} \neg \moveL(i,t) \vee \bigvee_{i\in L_0(v)}
    \moveL(i,t) \right.\\
    & \left. \vee \neg \timee(t) \vee \board(B,v,t-1) \vee \board(W,v,t) \mid  v
    \in V, t \in T_\white \vphantom{\bigvee_{v \in e}} \right\} \label{eq:whiteL}
  \end{split}
\end{align}

This clause can also be read as ``Either the game is over, the vertex is already occupied or white chooses a move corresponding to the respective bits''.

\paragraph{Restricting Black.}

The variables encoding Black's choice are existentially quantified.
For correctness, we need to encode that Black does not claim too many vertices.
In fact, we need to encode that there is at most one change in the board.
This simply results in the clauses \eqref{eq:blackL1} and \eqref{eq:blackL0}.
If we observe a change in a board configuration, the corresponding bit is forced.
The clauses can be seen as the reverse of clauses \eqref{eq:whiteL} for White.
However, note that these clauses do not need an additional $\timee$ variable because once the game is over, the board is fixed, and the clauses will be activated.

\begin{align}
\!\!\!& \{ \board(B,v,t\!-\!1) \!\vee \!\neg\board(B,v,t) \!\vee\! \moveL(i,t)     \mid  v \in V, t \in T_\black, i \in L_1(v)\} \label{eq:blackL1} \\
\!\!\!& \{ \board(B,v,t\!-\!1) \!\vee \!\neg\board(B,v,t) \!\vee\! \neg\moveL(i,t) \mid  v \!\in V, t \!\in T_\black, i \!\in L_0(v)\} \label{eq:blackL0}
\end{align}

\paragraph{Winning configuration.}

Black is required to win the game for the formula to be evaluated to true.
White should not have been able to steal a winning configuration before Black won.
The basic functionality of the encoding is to propagate the board forward to the final time point (see implication chain \eqref{eq:enc14-4}).
Only at the last time point $F$ it is checked if black or white has won.
For each winning configuration $e\in E_\black$, a variable $\win(e)$ was introduced to encode the winning configurations.
Clauses \eqref{eq:enc5-11a} encode that at least one of the winning configurations has to be reached, and \eqref{eq:enc5-11b} defines which vertices belong to the hyperedge.

White should never reach one of its winning positions.
This is enforced by clause \eqref{eq:enc5-10}, for each hyperedge, white should never own all vertices.
This looks straightforward from the definition, but we need to make sure with other clauses that White cannot play illegal moves to reach a winning position.
\begin{align}
    \left\{\bigvee_{e\in E_\black} \win(e) \right\} &\label{eq:enc5-11a} \\
    \left\{\neg \win(e) \vee \board(\black,v,\depth) \right. & \left. \mid v \in e, e \in E_\black
    \right\}\label{eq:enc5-11b} \\
   \left\{ \bigvee_{v \in e} \neg \board(\white,v,\depth) \right. & \left.  \relmiddle{|} \vphantom{\bigvee_{v \in e}} e \in E_\white \right\}\label{eq:enc5-10}   \\
    \left\{\win(e) \vee \bigvee_{v \in e} \neg \board(\black,v,\depth) \right. & \left.  \relmiddle{|} \vphantom{\bigvee_{v \in e}} e \in E_\black \right\}\label{eq:enc5-10x}
\end{align}

\subsection{Extensions}

To keep the description of the clauses clean and clear, we only describe the
following simple extensions to \ENCP.

\emph{Initial positions.} There is a simple way to translate positional games
that contain initial positions of White and Black, i.e., vertices that players
own before the actual game starts.
Rewrite the game into an equivalent description without initial positions: for each initial position $v$ of one player, remove this vertex from all its winning configurations and all winning configurations of the opposing player that contain $v$.
After this operation, we can remove $v$ from $V$ and have an equivalent game.

\emph{Symmetry breaking: Initial Moves.}
We employ a simple form of manual symmetry breaking by restricting the set of
vertices from which the first move can be chosen.
For instance, in \ac{GTTT} and a $n\times n$ board, if this set contains the upper left triangle of the board (the set of coordinates $(i, j)$ such that $1 \leq i \leq j \leq n/2$), the symmetries of the squared board are broken.
Typically, for other games with some initial positions for White and Black,
there is not much need for symmetry breaking since row or column symmetries are
usually already broken by such a position. 

\emph{Symmetry breaking: Consecutive moves.}
The positional game description need not have Black and White alternate moves:
a description may allow a player to select several vertices consecutively.
For instance, when $q=2$, players claim two vertices in each round.
To simplify the presentation of \ENCP we have not introduced clauses that break this symmetry.

\section{Comparison to Related Encodings}\label{related-works}


While the \ENCP Encoding we have developed can be described succinctly and looks like a rather direct implementation of the rules of positional games, one might even call it straightforward, this is the result of a long refinement process.
Through several design iterations, we have obtained an encoding that we judge elegant and that should provide a flexible starting point for future explorations in the encodings of games.
As we shall see in Section~\ref{sec:analysis}, this apparent simplicity does not come at a computational price, on the contrary.

\emph{Adapted Log Encoding.}
One of the key insights that contribute to the elegance and performance of \ENCP is that we can avoid so-called \emph{cheating variables} through a logarithmic encoding of the players' choice.
We devised this technique spontaneously, but we later realized that it had been described and advocated before our work.
To the best of our knowledge, this concept was first introduced for the encoding of quantified constraint satisfaction problems (CSPs) into \ac{QBF} \cite{Gent042}.
This technique was also applied in a game encoding to \ac{QBF} \cite{Gomes05}.
The basic idea: a logarithmic number of universal variables encode the binary representation of a CSP variable.
The motivation in this related work was the same as ours:, namely to avoid auxiliary variables that represent cheating and make the overall encoding more complicated.

\subsection{Comparison to \DYS}

The closest to our encoding is the \DYS encoding of \ac{GTTT}
\cite{DiptaramaYS2016} that has been proposed recently, which is an adaptation
of the encoding in \cite{GentA2003}. The structure and clauses for these two
encodings are similar, so our more detailed comparison to \DYS also applies to
the encoding in \cite{GentA2003}.

Apart from these games and encodings, to the best of our knowledge, we are not
aware of any other QBF formulations of games that improve upon them. In the
remainder of this section, we will go into various properties regarding \ENCP and
the existing encodings.

\emph{Generalization.}
Although we presume that the ideas behind \DYS could be extended to arbitrary
positional games, the description of the encoding was tailored to the \ac{GTTT}
domain. We chose positional games as an input formalism to reach a reasonable
level of generalization, i.e., many two-player can be formulated as positional
games, but these domains share enough structural properties from the description to create neat
encodings.

\emph{Clausal Description.} The description of the clauses in \DYS is not
purely clausal and contains many equivalences. The general translation of
equivalences into CNF introduces auxiliary (Tseytin) variables of which some
can be avoided through better techniques. Our description consists only of
clauses, and much work has gone to reduce the number of variables.


\emph{Binary Clauses.} Binary clauses enjoy many theoretical and practical
advantages. A purely 2CNF problem can be solved in polynomial time, SAT solvers
invest in the special treatment of binary clauses to speed up propagation and
learning. This should also apply to \emph{QBF} solving. Discovering a binary
clause structure of a certain aspect of a problem description might be the key to
crafting encodings that also solve fast. Our encoding demonstrates that many
aspects of positional games can be captured through sets of binary clauses
forming implication chains. 

\emph{Timing.} The variable $gameover_z$ in \DYS  has the same meaning as
$\timee$ in our encoding, it marks the end of the game and is crucial to
prevent white from reaching a winning position after black has already won. In
\DYS this variable is added to almost all clauses such that, when the game is
finished, the universal variables cannot falsify these clauses anymore. This
technique produces correct encodings, but affects propagation negatively and
weakens clause learning. 
We avoid such weakening of the other clauses by de-coupling time and the board
structure of a game. When the game is finished (i.e.  $\timee$ is set to false)
independently of the choices of the players, no moves are allowed anymore and
all empty board positions are propagated through to the end. We expect that
\ENCP makes it easier for the solver to exploit the transposition of a sequence of
moves leading to the same board configuration as previous encodings.

\emph{Monotonicity.} Using the property of monotonicity of positional
games---the set of claimed vertices only grows throughout the game---our
clauses manage to capture this property more directly than the previous
encodings. For instance, \DYS introduces variables that are true if a player
wins in time point $t$ by reaching a winning configuration.  We avoid the need
to know explicitly by which time point a player won via propagating the claimed
vertices through to the last time point and only need to test for the winning
configuration of both players there. Through de-coupling the time aspect of
games from checking winning positions, our encoding has fewer variables and
shorter clauses, that again benefit propagation. 

\emph{Indicator Variables.} 
To the best of our knowledge, all translations of games (including
non-positional ones) to \ac{QBF} introduce variables that indicate types of
illegal moves by White. 
Such variables again weaken the encoding due to longer clauses. 
The encoding \DYS has the following types of illegal moves; white claims too many vertices, too few vertices, already occupied vertices. 
All of these are explicitly encoded in \DYS, whereas \ENCP corrects white's move. 
The key insight in the work of \cite{Gomes05} raises the question of how to
address White's illegal behavior without weakening the encoding or introducing
auxiliary variables.

\subsection{Size of the \ENCP Encoding}\label{sec:size}

In Table~\ref{tab:readability}, we give a high-level overview of the different variables and clauses of the two encodings \DYS and \ENCP.
We rename the variables from \DYS for clarity to the nearest corresponding variables in \ENCP.
We believe that an encoding having fewer different types of variables and with a more uniform set of clauses is easier to understand and can provide easier inspiration to future game encodings.

\begin{table}
\caption{Compared readability of the \ENCP and \DYS encodings.}
\label{tab:readability}
\centering
\begin{tabular}{c@{\quad}c@{\quad}ccc}
\toprule
& \DYS & \ENCP \\
\midrule
variables & 9: $\left\{\begin{array}{l}\text{time}, \text{board}, \text{occupied}, \text{win},\\ \text{wins},\text{move}, \text{cheat}, \text{cheats}\end{array}\right\}$ & 4: $\left\{\begin{array}{l}\text{time}, \text{board},\\ \text{win}, \text{move}\end{array}\right\}$\\
formulas & 23: Equations (non clausal) & 13: Clause types \eqref{eq:enc14-1}--\eqref{eq:enc5-10x} \\
\bottomrule
\end{tabular}
\end{table}

It is straightforward to estimate the number of variables and clauses of \ENCP.
For instance, clauses with a description containing $v\in V, t\in T$ are generated at most $\size\depth$ times, where the size of the board is $\size=|V|$ and the length of the game $\depth=|T|$.
The encoding \DYS in \cite{DiptaramaYS2016} is originally formulated only for generalized \gamename{Tic-tac-toe}.
For a fair comparison, we estimate the number of clauses of \DYS and choose advantageous translation from non-clausal formulas.

Table~\ref{tab:size} contains a detailed list of variables and clauses introduced by \ENCP (compared with \DYS).
In the table, we differentiate between binary, ternary clauses (two, respectively three literals) and longer clauses.
Generally, short clauses are to be preferred because they generalize more and tend to provide more optimization opportunities to solvers.

\begin{table}
\caption{Compared succinctness of \ENCP and \DYS in terms of number of variables and clauses.}
\label{tab:size}
\centering
\begin{tabular}{@{}cc@{\quad}c@{\quad}ccc@{}}
\toprule
    & \multicolumn{2}{c}{variables} & \multicolumn{3}{c}{clauses} \\
\cmidrule(lr){2-3} \cmidrule(lr){4-6}
    & $\exists$  & $\forall$ & binary & ternary & long\\
\midrule
\DYS& $13\size\depth$ & $\size\depth$ & $ 3 \depth$ & $14 \size \depth + 4 \size^2\depth$ & $10\size \depth + |E_\black| \size\depth + |E_\white|\depth$ \\
 \ENCP   & $2\size\depth+\lg(\size)\depth$ & $\lg(\size)\depth$ & $ 3 \size\depth + \size |E_\black|$ & $4 \size \depth + \size\lg(\size)\depth$ & $\size \depth + |E_\black| + |E_\white|$ \\
\bottomrule
\end{tabular}
\end{table}

\section{Instances}
\label{sec:instances}

We used the encoding above to generate three sets of \ac{QBF} instances based on some well-known positional games.\footnote{All our generated instances are available at \\ \url{github.com/vale1410/positional-games-qbf-encoding}.}
The first two sets consist of positions of \gamename{Hex} and of a generalization of \gamename{Tic-tac-toe} on boards that are relatively small by human playing standards.
Positions from that benchmark are fairly easy to solve even for relatively inexperienced human players of these games, and they can be solved almost instantaneously by specialized solvers.
Our encoding of these positions should provide a reasonable challenge for \ac{QBF} solvers as of 2019.

The third set contains the starting position of 4 positional games that are of interest to experienced human players and mathematicians.
At least 3 of these positions can be solved by specialized game algorithms developed in the 1990s and 2000s albeit with a non-trivial programming effort.
The instances in this third set are out of reach of current \ac{QBF} solvers and we believe that solving these positions with a \ac{QBF} solver---via our encoding or a better one---can constitute a good milestone for the field.

\subsection{Harary's Tic-tac-toe and \ac{GTTT}$(p, q)$}
\label{sec:harary}

\gamename{Harary's Tic-tac-toe} is a Maker-Maker generalization of \gamename{Tic-tac-toe} where instead of marking three aligned stones, the players are trying to mark a set of cells congruent to a given polyomino.
This type of game has received accrued interest from the mathematical community, which was able to show the existence of a winning strategy or lack thereof for most polyomino shapes.
\acs{GTTT}$(p, q)$ is a further generalization of \gamename{Harary's Tic-tac-toe} along the principle of $(p, q)$-positional games~\cite{DiptaramaYS2016}.

Previous work has already proposed an encoding of \acs{GTTT}$(p, q)$ played on small boards to \ac{QBF}~\cite{DiptaramaYS2016}.
We refer to this existing in encoding as \emph{DYS}.
In our first set of benchmarks, we encode the exact same \acs{GTTT}$(p, q)$ configurations as in previous work.
However, since our encoding is different, with obtain a different set of \ac{QBF} instances.
This provides us with an opportunity to directly compare our approach with existing work.
We report results on the 96 instances of \acs{GTTT}$(1, 1)$ played on a $4\times 4$ board and compare formula size and solving performance with the DYS encoding.\footnote{We also generated the instances for larger values of $p$ and $q$ but the formulas are much easier to solve and provide less insight.}


\subsection{Hex}
\label{sec:hex-instances}

We use 20 hand-crafted \gamename{Hex} puzzles of board size $4 \times 4$ up to $7\times 7$ that all have a winning strategy for Player 1.
The first 19 of these \gamename{Hex} instances are of historical significance.
Indeed, they were created by Piet Hein, one of the inventors of \gamename{Hex} and first appeared in the Danish newspaper \emph{Politken}~\cite{Hein1942-1943,Maarup2005} during World War II~\cite{HaywardT2019}. 
The remaining puzzle is a $5 \times 5$ position proposed by Cameron Browne, it arose during standard play and offers a significant challenge for the \ac{MCTS} algorithm and the associated RAVE enhancement~\cite{Browne2013}.
This is noteworthy because \ac{MCTS} is the foundational algorithm behind the top artificial players for numerous games, including \gamename{Hex} and \gamename{Go}~\cite{BrownePWLCRTPSC2012}.


\subsection{Challenges}

We now put forward a few positional games that have attracted the attention of board game players as well as AI or mathematics researchers.
Table~\ref{tab:challenges} summarizes the proposed challenges together with the size of their \ENCP encoding.

\gamename{Qubic}, also known as \gamename{$3$-dimensional Tic-tac-toe}, is played on a $4 \times 4 \times 4$ cube, and the goal is to mark 4 aligned cells, horizontally, vertically, or diagonally.
Our first domain was solved for the first time in 1980 by combining depth-first search with expert domain knowledge~\cite{Patashnik1980}.
A second time in the 1990s using \ac{PNS}, a tree search algorithm for two-player games~\cite{vdHerikUvRan2002games}.

The second domain, \emph{freestyle} \gamename{Gomoku}, is played on a $15\times 15$ board and the goal is to mark 5 aligned cells, horizontally, vertically, or diagonally.
Already in the 1930s, \gamename{Gomoku} was perceived to be giving an overwhelming advantage to Black~\cite{vdHerikUvRan2002games}, the starting player, and by the 1980s professional \gamename{Gomoku} players from Japan had claimed that the initial position admitted a Black winning strategy~\cite{AllisvdHH1996}.
This was confirmed in 1993 using the \ac{PNS} algorithm, a domain heuristic used to dramatically reduce the branching factor, and a database decomposing the work in independent subtasks~\cite{AllisvdHH1996}.

\gamename{Connect6} is akin to \gamename{Gomoku}, but the board is $19\times 19$, the goal is 6 aligned cells, and players place two stones per move~\cite{HsiehT2007}.
The \emph{Mickey Mouse} setup once was among the most popular openings of \gamename{Connect6} until it was solved in 2010~\cite{WuLSKLCC2013}.
The resolution of \gamename{Connect6} was based on \ac{PNS} distributed over a cluster.

Our last challenge is an open-problem in \gamename{Harary's Tic-tac-toe} which corresponds to achieving shape \gamename{Snaky} \begin{tikzpicture}[scale=0.15]
    \draw[thick] (1,0) rectangle ++(1,1);
    \draw[thick] (2,0) rectangle ++(1,1);
    \draw[thick] (3,0) rectangle ++(1,1);
    \draw[thick] (4,0) rectangle ++(1,1);
    \draw[thick] (4,1) rectangle ++(1,1);
    \draw[thick] (5,1) rectangle ++(1,1);
\end{tikzpicture} on a $9\times 9$ board.
This problem was recently put forward as an intriguing challenge for \ac{QBF} solvers~\cite{DiptaramaYS2016} and we offer here an alternative, more compact, encoding.



\begin{table}
    \centering
    \caption{Selected problems put forward to the \ac{QBF} community and size
    of their \EncodingNameP encoding. No preprocessing has been applied to these
    instances.}
    \label{tab:challenges}
    \begin{tabular}{llcrrrrr}
        \toprule
        \multicolumn{2}{l}{Challenge problem} & \multirow{2}{*}{\parbox{12mm}{First systematic solution}}  & \multicolumn{5}{c}{Size in \acs{QBF}} \\
        \cmidrule(r){1-2} \cmidrule(l){4-8}
        Domain & Variant & &  \#qb & \#$\forall$ & \#$\exists$ & \#cl & \#lits  \\
        \midrule
        \gamename{Qubic}    &   $4 \times 4 \times 4$               &   1980  &  63  &  186   & 4.5k   &  25.0k    & 102k  \\
        \gamename{Snaky}    &   $9 \times 9$                        &   open  &  81  &  280   & 7.5k   &  45.7k    & 188k  \\
        \gamename{Gomoku}   &   $15\times 15$ freestyle             &   1993  &  225 &  896   & 52.9k  &  360k   & 1514k \\
        \gamename{Connect6} &   $19\times 19$ Mickey Mouse          &   2010  &  179 &  1602  & 130k &  1020k  & 3070k \\
        \bottomrule
    \end{tabular}
\end{table}

\section{Analysis}
\label{sec:analysis}

\subsection{Setup of Experiments}

When solving problems encoded in \ac{QBF}, the ideas underlying the encoding of
a problem is only a factor in whether the instances can be solved within
reasonable resources.
Two other important factors are the specific solver invoked and the preprocessing performed on the instance before solving, if any.
In our experiments, we chose four state-of-the-art QBF solvers, including the top three solvers of the 2019 QBF Competition\footnote{\url{http://www.qbflib.org/eval19.html}} and three preprocessors, as indicated in Table~\ref{tab:software}.\footnote{We also attempted to use the \texttt{rareqs} \ac{QBF} solver, but it timed out on almost all instances.
  Preprocessor H was omitted due to large timeouts.}
All software was called with default command line parameters.

\begin{table}
\caption{Solvers and preprocessors used in the experiments.}
\label{tab:software}
\centering
\begin{tabular}{@{}lll@{}}
\toprule
& Software & Shorthand\\
\midrule
\multirow{3}{*}{\rotatebox{90}{Solver}}
& DepQbf 6.03 \cite{depqbfLonsing} & depqbf \\
& Caqe   4.0.1 \cite{caqeTentrup} & caqe   \\
& Qesto  1.0  \cite{qestoJanota}  & qesto  \\
& Qute   1.1  \cite{qute17}  & qute   \\
\midrule
\multirow{4}{*}{\rotatebox{90}{\shortstack{Prepro\\cessor}}}
& QratPre+ 2.0 \cite{LonsingE2019} & Q \\
& HQSPRE   1.4 \cite{hqspre+} & H \\
& Bloqqer  v37 \cite{bloqqer} & B \\
& None & N \\
\bottomrule
\end{tabular}
\end{table}
\begin{table}
    \caption{Solver performance on the first benchmark depends on the encoding, always using the best preprocessor. Timeout 1000s.}
\label{tab:gttt44-speed}
\centering
\begin{tabular}{@{}lll*{5}{r}}
\toprule
    & Solver & Preproc. & S & $\top$ & $\bot$ & U & time(s) \\
\midrule
\multirow{4}{*}{\DYS}
    & caqe    & B  & 92  & 31  & 61  &  4  & \bftab 11468 \\
    & depqbf  & Q  & 82  & 28  & 54  & 14  & 27211 \\
    & qesto   & BQ & 74  & 27  & 47  & 22  & 34887 \\
    & qute    & BQ & 69  & 27  & 42  & 27  & 35837 \\
\midrule
\multirow{4}{*}{\ENCP}
    & caqe    & B  & 96  & 34  & 62  &  0  & 568  \\
    & depqbf  & B  & 96  & 34  & 62  &  0  & 578   \\
    & qesto   & B  & 96  & 34  & 62  &  0  & \bftab 499 \\
    & qute    & B  & 96  & 34  & 62  &  0  & 8243 \\
\bottomrule
\end{tabular}
\end{table}

The experiments were run on an i7-7820X CPU @ 3.60GHz with 8 cores and
32GB RAM. It was made sure that each solver had a dedicated single core
available. No solver needed multi-threading.

\subsection{Experimental comparison of \ENCP to the \DYS}
Before attempting to solve positional games, let us first examine how large and amenable to preprocessing the generated encodings are.
We use an approach inspired by recent work on QBF preprocessors~\cite{LonsingE2019} and report in Table~\ref{tab:preproc} the number of quantifier blocks, universal and existential variables, clauses, and literals, as well the time needed for the preprocessing of a representative instance of GTTT(1, 1).
On both the existing \DYS encoding and our proposed \ENCP, we test each preprocessor individually and the outcome of running one preprocessor and then another.
No preprocessor timed out.

\begin{table}
\caption{Preprocessing on  $5\times 5$ instance \texttt{gttt\_1\_1\_00101121\_5x5\_b}}
\label{tab:preproc}
\centering
\begin{tabular}{ll@{\quad}*{8}{r}}
\toprule
    &  &  N      &  Q      &  H      &  B      &  QB     &  BQ     &  HQ     &  QH       \\
\midrule
\multirow{6}{*}{\DYS}
&  \#qb       &  25     &  25     &  25     &  25     &  25     &  25     &  25     &  25       \\
&  \#$\forall$&  300    &  300    &  300    &  299    &  299    &  299    &  300    &  300      \\
&  \#$\exists$&  21056  &  12058  &  7553   &  2750   &  2605   &  2750   &  7553   &  7545     \\
&  \#cl       &  53.6k  &  35.9k  &  33.0k  &  21.4k  &  19.6k  &  19.3k  &  30.2k  &  30.4k    \\
&  \#lits     &  191k   &  127k   &  145k   &  120k   &  107k &  106k &  103k &  135k   \\
&  time(s)  & 0      &  46     &  1210   &  9      &  55     &  22     &  1233   &  2030     \\
\midrule
\multirow{6}{*}{\ENCP}
&  \#qb       &    25    &  25     & 25    &  25    &  25    &  25    &  25    &  25     \\
&  \#$\forall$&    60    &  60     & 56    &  57    &  57    &  57    &  56    &  56     \\
&  \#$\exists$&    826   &  751    & 1973  &  1071  &  1083  &  1071  &  1973  &  1973   \\
&  \#cl       &    3.9k  &  3.0k   & 9.9k  &  5.6k  &  5.9k  &  5.4k  &  8.9k  &  9.3k   \\
&  \#lits     &    15k &  11k  & 35k &  23k &  26k &  23k &  25k &  34k  \\
&  time(s)    &     0    &   0     &  13   &   0    &   0    &   0    &   13   &   12    \\
\bottomrule
\end{tabular}
\end{table}

Two observations stand out when looking at Table~\ref{tab:preproc}.
First, \DYS is much larger than \ENCP across most of the size dimensions.
Second, preprocessors seem to be much more capable or reducing the size of the \DYS instance than the size of the \ENCP instance.
Our interpretation is that it is a direct consequence of the effort we have put in crafting the proposed new encoding: there is relatively little improvement room left for the preprocessors to improve the formulas.
Since the size of a formula directly impacts how hard it is to solve, we expect \ac{QBF} solvers to struggle much more with \DYS-encoded game instances than with \ENCP-encoded ones.

In our next experiment, we compare how well QBF solvers manage to solve GTTT game instances when encoded with \DYS and with \ENCP.
Since different preprocessors tend to play to the strength of different solvers, we report the preprocessor that leads to the best performance for each solver separately.
We compare the solvers and the encodings using 96 GTTT(1, 1) $4\times 4$ game instances and assuming a timeout of 1000s, Table~\ref{tab:gttt44-speed} displays for each configuration the number of formulas solved (S), proven satisfiable ($\top$), proven unsatisfiable ($\bot$), and unsolved (U), as well as the cumulative time spent by the solver.

The data in Table~\ref{tab:gttt44-speed} confirms our intuition.
GTTT games can be more effectively solved through our encoding: 3 out of 4 solvers solve all \ENCP instances, whereas none solve all \DYS instances, and \ENCP instances are solved between up to two orders of magnitude faster.
Furthermore, our results demonstrate that the choice of encoding has a bigger impact than the choice of solver and preprocessor.


\subsection{Solving increasingly realistic games}

In this Section, we investigate how well the combination of our new \ENCP Encoding and state-of-the-art \ac{QBF} solver fare when facing positional game instances that are realistically sized.
In other words, can a well-chosen combination of encoding and solver solve instances that to non-expert human players find challenging?
The answer is clear: yes, as we shall demonstrate through two qualitative examples and a quantitative experiment.

As a first example, consider Figure~\ref{fig:gtttDeep}.
It displays the record of a \ac{GTTT} $5\times 5$ game between the authors of this paper, where the objective was to create an L-shape.
As we can see, the game was won by Player Black after 19 moves were played.
At a certain stage in the game, both players recognized that the position was heavily favorable to Player Black and that with some effort, one might even be able to visualize a winning line.
We then wondered how close to the end of the game the solver could detect Black's winning strategy.
To answer this question, for each recorded position where Black is to play, we generated a range of \acp{QBF}, one per candidate depth of Black winning strategy.
We provide representative results in Figure~\ref{fig:gtttDeepTable}, where we display the critical depth $d$ that Black needs to win from the position.
For instance, once $k=10$ moves were played (5 by Black and 5 by White), our solver caqe-B-\ENCP needed less than 0.1s to establish that there is no winning strategy of depth $9$ and needed $0.2$ to identify that there was a winning strategy of depth $d=11$.

\begin{figure}
\newcommand{\mltr}[1]{\multirow{2}{*}{#1}}
\renewcommand{\scala}{0.7}
\subfloat[\label{fig:gtttDeep} The game after Black has claimed the L shaped winning position. ]{
\raisebox{-.55\height}{\begin{tikzpicture}[>=stealth',scale=\scala,every node/.style={scale=2.*\scala}]
\draw[thick, brown, step=1.0cm,xshift=-0.5cm, yshift=-0.5cm] (1,1) grid +(5,5);
\draw[ultra thick] (1.5, 1.5) -- (1.5, 5.5) -- (2.5, 5.5) -- (2.5, 2.5) -- (3.5, 2.5) -- (3.5, 1.5) -- (1.5, 1.5);
\node[hav-black] at (2, 2) {1};
\node[hav-white] at (1, 1) {2};
\node[hav-black] at (3, 2) {3};
\node[hav-white] at (4, 2) {4};
\node[hav-black] at (3, 3) {5};
\node[hav-white] at (3, 1) {6};
\node[hav-black] at (3, 4) {7};
\node[hav-white] at (3, 5) {8};
\node[hav-black] at (2, 4) {9};
\node[hav-white] at (2, 1) {10};
\node[hav-black] at (4, 1) {11};
\node[hav-white] at (1, 2) {12};
\node[hav-black] at (4, 4) {13};
\node[hav-white] at (1, 4) {14};
\node[hav-black] at (1, 3) {15};
\node[hav-white] at (5, 4) {16};
\node[hav-black] at (2, 3) {17};
\node[hav-white] at (4, 3) {18};
\node[hav-black] at (2, 5) {19};
\end{tikzpicture}}
}
\hfill
 \subfloat[\label{fig:gtttDeepTable}Time (s) needed by caqe-B-\ENCP to establish
     whether Black has a winning strategy of depth $d$ and $d-2$ in the position
     obtained after $k=0\ldots12$ stones have been placed.]{
 \begin{tabular}{@{}rr*2{@{\qquad}r@{.}l}@{}}
   \toprule
  $k$ & $d$ & \mltc{$\not\models \phi_{d-2}^k$} & \mltc{$\models \phi_{d}^k$} \\
  \midrule
0  &  15(?) & 1088& &  \mltc{$>$ 8 hours} \\
2  &  15(?) & 2605& &  \mltc{$>$ 8 hours} \\
4  &  13    & 67&  & 147& \\
6  &  13    & 10&6 & 101& \\
8  &  11    & 1&6 & 3&6 \\
10 &   9    &  0&06 & 0&2 \\
12 &   7    &  0&03 & 0&06 \\
\bottomrule
\end{tabular}
}
\caption{Game 1 on $5\times 5$ with shape L where Black has an established long-term winning strategy from move 4 onwards.}
\label{fig:idqbf1}
\end{figure}

Caqe-B-\ENCP cannot establish the status of the game within the first two moves, except for proving that there is no Black winning strategy of 13 moves or less.
However, it is remarkable that within 3 minutes of thinking time (147 seconds to be precise), Caqe-B-\ENCP can demonstrate that after the second White move, Black had a sure win in 13 moves.
We may also observe that the human player was not as sharp since their Black move 5 did not progress towards winning the game, and the shortest winning strategy after move 6 is also 13 moves.

Our next example, Figure~\ref{fig:idqbf2}, involves the same positional game but a different initial sequence of moves such that Black has a fairly short winning strategy from move 5 already.
Our analysis takes a different turn here as we fix the position 5-moves in and ask how long Caqe-B-\ENCP needs to establish that there exists a winning strategy of various depths.
Since any $d$-move winning strategy is a special case of a $d+2$-move winning strategy, a human reasoner typically does not require more time to establish that Black can win within 15 moves than within 5 moves.

\begin{figure}
\newcommand{\mltr}[1]{\multirow{2}{*}{#1}}
\renewcommand{\scala}{0.7}
\subfloat[\label{fig:gttt55id}Position after White's mistaken second move.]{
\raisebox{-.55\height}{\begin{tikzpicture}[>=stealth',scale=\scala,every node/.style={scale=2.*\scala}]
\draw[thick, brown, step=1.0cm,xshift=-0.5cm, yshift=-0.5cm] (1,1) grid +(5,5);
\node[hav-black] at (3, 3) {1};
\node[hav-white] at (1, 1) {2};
\node[hav-black] at (2, 3) {3};
\node[hav-white] at (2, 1) {4};
\end{tikzpicture}}
}
\hfill
\subfloat[Time (s) needed by caqe-B-\ENCP to establish whether Black has a winning strategy of depth $\leq d$ in the position in Figure~\ref{fig:gttt55id}.]{
\hspace{10mm}
\begin{tabular}{r@{\quad}*2{r@{.}l@{\qquad}}r@{.}l}
\toprule
$d$ &\mltc{$\not\models\phi_d$}&\mltc{$\models\phi_d$} \\
\midrule
 1   & 0&01 & \mltc{}\\
 3   & 0&01 & \mltc{}\\
 5   & \mltc{} & 0&05   \\
 7   & \mltc{} & 0&12   \\
 9   & \mltc{} & 4&08   \\
 11  & \mltc{} & 13&  \\
 13  & \mltc{} & 113& \\
 15  & \mltc{} & 1192& \\
\bottomrule
\end{tabular}
\hspace{10mm}
}
\caption{GTTT $5\times 5$ L game where after 4 moves, Black can force a 5-move win.}
\label{fig:idqbf2}
\end{figure}

Unsurprisingly, things are harder for the artificial reasoner in its basic form, as seen from Figure~\ref{fig:idqbf2}.
Indeed, iterative deepening, an algorithmic principle in-game search, recommends searching for a shallow winning strategy before attempting to find a deeper one.
This principle lets one benefit from the memory efficiency of depth-first search and the completeness of breadth-first search.
It is easily adapted to solving games via \ac{QBF}: encode one formula per depth and attempt to solve them one-by-one in order.
We demonstrate the benefits of this adaptation in Figure~\ref{fig:idqbf2}: the position admits a depth $5$ winning strategy.
Proving the existence of a strategy of depth $\leq 5$ needs less than $0.1$ second, but the formula stating the existence of a strategy of depth $\leq 15$ needs close to 20min to be proven.
Although the outcome of searches at short depths is subsumed by that of deeper searches, the exponential growth of the required solving time makes iterative deepening a worthy trade-off.

Our final benchmark is the set of 20 historical \gamename{Hex} puzzles
described in Section~\ref{sec:hex-instances}. All positions on board sizes
$5\times 5$ or less can be solved by state-of-the-art \ac{QBF} solvers
(Table~\ref{tab:hex}).
From the $6\times6$ puzzles, one is solvable within the timeout of 8h, and for the others two solvers can show a lower bound of depth $13$.
The $7\times7$ puzzle remains out of reach at this stage.
This is a remarkable feat: for the first time, the \ac{QBF} technology can address game situations considered of interest to human players.

\begin{table}
\caption{Solving classic \gamename{Hex} puzzles by encoding them through \ENCP. Timeout 8h.}
\label{tab:hex}
\centering
\begin{tabular}{lcc*6{r@{.}l}rr}
\toprule
Puzzle & size & depth & \multicolumn{4}{c}{caqe-B}  & \multicolumn{4}{c}{depqbf-BQ}  & \multicolumn{4}{c}{qesto-B}  \\
& & $d$ & \mltc{$\not\models \phi_{d-2}$} & \mltc{$\models \phi_{d}$} & \mltc{$\not\models \phi_{d-2}$} & \mltc{$\models \phi_{d}$} & \mltc{$\not\models \phi_{d-2}$} & \mltc{$\models \phi_{d}$} \\
\midrule
Browne  & 5x5 & 09 &   0&11 &  1&38   &     0&23   & 6&89     &    0&07  &  1&35    \\      
Hein 04 & 3x3 & 05 &   0&01 &  0&01   &     0&00   & 0&00     &    0&00  &  0&00    \\      
Hein 09 & 4x4 & 07 &   0&03 &  0&07   &     0&01   & 0&07     &    0&01  &  0&08    \\      
Hein 12 & 4x4 & 07 &   0&01 &  0&06   &     0&00   & 0&03     &    0&01  &  0&02    \\      
Hein 07 & 4x4 & 09 &   0&09 &  0&62   &     0&09   & 0&88     &    0&06  &  0&44    \\      
Hein 06 & 4x4 & 13 &   0&56 &  1&43   &     0&61   & 6&07     &    0&27  &  0&80    \\      
Hein 13 & 5x5 & 09 &   0&09 &  2&51   &     0&15   & 3&79     &    0&04  &  0&58    \\      
Hein 14 & 5x5 & 09 &   0&10 &  0&65   &     0&18   & 3&39     &    0&09  &  1&16    \\      
Hein 11 & 5x5 & 11 &   0&52 &  2&36   &     2&01   & 24&54    &    0&61  &  5&21    \\      
Hein 19 & 5x5 & 11 &   0&15 &  1&78   &     0&26   & 3&02     &    0&12  &  0&78    \\      
Hein 08 & 5x5 & 11 &   1&16 &  6&17   &     1&01   & 17&20    &    0&26  &  1&97    \\      
Hein 10 & 5x5 & 13 &   3&48 &  25&00  &     23&94  & 334&30   &    7&66  &  11&44   \\      
Hein 16 & 5x5 & 13 &   7&09 &  47&57  &     99&54  & 672&14   &    12&76 &  91&45   \\      
Hein 02 & 5x5 & 13 &   27&24&  27&23  &     231&97 & 871&44   &    29&33 &  127&86  \\      
Hein 15 & 5x5 & 15 &   42&24&  21&67  &     62&27  & 391&57   &    7&50  &  15&85   \\      
Hein 05 & 6x6 & 13    &   1064&  &  7734&     &  \mltc{TO} & \mltc{TO} & 196&       & 1345&   \\      
Hein 17 & 6x6 & 15(?) &   5700&  &  \mltc{TO} &  \mltc{TO} & \mltc{TO} & 3682&      & \mltc{MO}  \\      
Hein 03 & 6x6 & 15(?) &   3998&  &  \mltc{TO} &  \mltc{TO} & \mltc{TO} & 2792&      & \mltc{MO}  \\      
Hein 20 & 6x6 & 15(?) &   21024& &  \mltc{TO} &  \mltc{TO} & \mltc{TO} & 17019&     & \mltc{MO}   \\      
\bottomrule
\end{tabular}
\end{table}

\section{Conclusion and Future Work}

We consider the craft of finding efficient translations of a problem description to the clausal representation an important step towards better performances of \ac{QBF} solvers.
There is an extensive body of work analyzing CNF encodings for SAT solving.
Much less is known about \ac{QBF} and how different CNF representations of constraints affect solving times.
Our investigation regarding the class of positional games demonstrates that a carefully crafted translation using structural properties to decrease and shorten clauses and decrease the number of variables improves the applicability beyond trivial problems.
We list some key insights.

\begin{itemize}
\item Binary implication chains capturing monotone structural properties of the problem are crucial.
\item Variables representing illegal moves by the universal player can be avoided.
\item Encoding the player choices logarithmically helps in this problem description.
\item Preprocessing is crucial for previous encodings to perform well, whereas it has a minor impact on our encoding.
\item QBF solvers are very sensitive to redundant variables.
\end{itemize}

Our investigation focused on clause representation and we have yet to extend to non-clausal description languages to \ac{QBF} such as QCIR.

The insights from our investigation can be applied to the translation of other almost-positional games and to planning problems with similar structures.
To the best of our knowledge, the challenge to solve \gamename{Connect4} via QBF (or a related logical reasoning method) with the full board of size $7\times 6$ remains open.




As a general technique, the translator could analyse the set of winning positions and generate a more compact clausal representation, for instance by introducing variables for common sub-structures.
Binary Decision Diagrams could help in this process~\cite{Edelkamp11}.

Although \gamename{Hex} is a positional game, its hypergraph representation is exponential in the board size because it needs to account for all paths between a pair of sides.
For larger boards one will need an implicit representation of the paths between the two sides, possibly drawing inspiration from existing first-order logic modeling~\cite{BonnetJS2016}.
Thus, a dedicated \ac{QBF} encoding treating reachability is necessary to deal with medium and large board sizes.
Exploring the design tradeoffs of such an encoding is a natural avenue for future work.

\bibliographystyle{plain}
\bibliography{main}

\end{document}